%% file: main.tex
\documentclass[acmtog,authorversion]{acmart}
\AtBeginDocument{%
  \providecommand\BibTeX{{%
    \normalfont B\kern-0.5em{\scshape i\kern-0.25em b}\kern-0.8em\TeX}}}

\copyrightyear{2024}
\acmYear{2024}
\setcopyright{rightsretained}
\acmConference[Authorversion]{}{2024}{}
\acmISBN{978-1-4503-XXXX-X/X/XX}

\usepackage{color, colortbl}
\usepackage{subfig}
\usepackage{multirow}
\usepackage{longtable}
\usepackage{hyperref}
\usepackage[font=small,labelfont=bf]{caption}
\usepackage{wasysym}
\usepackage[nolist,nohyperlinks, printonlyused, withpage]{acronym}
\usepackage{enumitem}

\begin{document}

\title[The Intricate Relationship Between Cognitive Biases and Dark Patterns]{Hell is Paved with Good Intentions: The Intricate Relationship Between Cognitive Biases and Dark Patterns}

\author{Thomas Mildner}
\email{mildner@uni-bremen.de}
\orcid{0000-0002-1712-0741}
\affiliation{%
  \institution{University of Bremen}
  \city{Bremen}
  \country{Germany}
  \postcode{28215}
}

\author{Albert Inkoom}
\email{ainkoom@uni-bremen.de}
\orcid{1234-5678-9012}
\affiliation{
    \institution{University of Bremen}
    \country{Germany}
}

\author{Rainer Malaka}
\email{malaka@uni-bremen.de}
\orcid{0000-0001-6463-4828}
\affiliation{
    \institution{University of Bremen}
    \country{Germany}
}

\author{Jasmin Niess}
\email{jasminni@ifi.uio.no}
\orcid{0000-0003-3529-0653}
\affiliation{%
  \institution{University of Oslo}
  \city{Oslo}
  \country{Norway}
  \postcode{}
}

\renewcommand{\shortauthors}{Mildner, Inkoom, Malaka, \& Niess}

\begin{acronym}
\acro{HCI}{Human-Computer Interaction}
\end{acronym}

\begin{abstract}
%Thomas
%Effective design leverages human cognition and perception to convey affordances, placing designers responsible for upholding users' ability to make informed decisions and avoid misleading interactions. However, this community has made a noticeable effort to record deceptive design patterns across various domains, revealing challenges for users in recognising and avoiding them. This research adds empirical insights to the discourse by exploring the intricate interplay between cognitive biases and deceptive design patterns. We conducted four focus groups with experts ($15$) in psychology and deceptive design, applying reflexive thematic analysis to learn about their thoughts on similarities, differences, and facilitators of the two subjects. Based on the results, we introduce the ``Relationship Model of Cognitive Biases and Deceptive Design Patterns''. The model emphasises ethical considerations for the design of interactions by fostering users' autonomy to make informed decisions. It further promotes reflection of deployed strategies by understanding the implied nuances of consequences.

Throughout the past decade, research in HCI has identified numerous instances of dark patterns in digital interfaces. These efforts have led to a well-fostered typology describing harmful strategies users struggle to navigate~\cite{gray_ontology_2023}. However, an in-depth understanding of the underlying mechanisms that deceive, coerce, or manipulate users is missing. We explore the interplay between cognitive biases and dark patterns to address this gap. To that end, we conducted four focus groups with experts ($N=15$) in psychology and dark pattern scholarship, inquiring how they conceptualise the relation between cognitive biases and dark patterns. Based on our results, we constructed the ``Relationship Model of Cognitive Biases and Dark Patterns'' which illustrates how cognitive bias and deceptive design patterns relate and identifies opportune moments for ethical reconsideration and user protection mechanisms. Our insights contribute to the current discourse by emphasising ethical design decisions and their implications in the field of HCI.

\end{abstract}



\maketitle

\textcolor{red}{\textbf{Draft: May 12, 2024}}

%%%%%%%%%%%%%%%%%%%%%%%%%%%%%
%       INTRODUCITON        %
%%%%%%%%%%%%%%%%%%%%%%%%%%%%%
\section{Introduction}
% dark patterns, unethical design has been linked to cognitive biases
The last decade of research in \ac{HCI} has shown a growing interest in unethical design practices --- in the midst of it are deceptive design practices and so-called ``dark patterns''~\footnote{We opted for the term ``dark pattern'' in alignment with previous authors when referring to identified harmful design practices. However, we are aware that the ACM Diversity, Equity, and Inclusion Council recently decided to classify the term as problematic, given its association with negative connotations. We use the term in line with its original context of \textit{hidden} consequences for users~\cite{brignull2015dark} and acknowledge that there currently exists no alternative terms that convey the full spectrum of deceptive, manipulative, obstructive, or coercive characteristics of the original term.}. Various streams of research have described instances that deceive or manipulate users in domains such as, but not limited to, e-commerce~\cite{mathur_dark_2019}, social media~\cite{mildner_about_2023, schaffner_understanding_2022}, and web and mobile interfaces~\cite{gray_dark_2018, di_geronimo_ui_2020, gunawan_comparative_2021}. While this body of work makes significant contributions that inform the protection of people against the harms embedded in these environments, we currently lack a fundamental understanding of the underlying principles that enable the deceptive, coercive, and potentially manipulative characteristics of dark patterns. Previous work by Mathur et al.~\cite{mathur_dark_2019} and Waldman~\cite{waldman_cognitive_2020} draw tentative connections to the exploitation of cognitive biases in this regard. However, the lack of supportive research leaves a gap for additional, fundamental work to describe this relationship in more detail. Furthermore, it remains unclear how this relationship could manifest itself and how it could be navigated in interaction design.

Generally, exploiting cognitive biases and connected heuristics~\cite{kahneman_perspective_2003,azzopardi_cognitive_2021} to manipulate users of any kind poses important ethical questions that require careful evaluation. Incorporating knowledge of human cognition and perception is an integral element of human factors and the design of interfaces. However, Thaler and Sunsteins' `nudge' theory~\cite{thaler_nudge_2008} and the principles of persuasive design~\cite{fogg_persuasive_2009} illustrate the effectiveness in which cognitive biases can be utilised to alter peoples' choice architecture; and, thus, require responsibility in designers not to hinder informed decision-making as users can be unaware of consequences~\cite{chen_practitioners_2022}. Wary of misuse, critique voices ethical concerns against paternalistic implications on agency~\cite{rizzo_little_2009, hausman_debate_2010, mccrudden_dark_2015}. Although Thaler and Sunstein independently addressed these concerns~\cite{thaler_nudge_2018,sunstein_nudges_2015} by reiterating how nudges were meant to empower people to make individual good decisions, Hansen and Jespersen~\cite{hansen_nudge_2013} argued for the necessity of nudging transparently upholding user agency.

This ongoing discourse spotlights the various ethical caveats designers should consider to support the informed decision-making of end-users. While recent work shows positive effects of design friction to assist users in making informed decisions~\cite{leimstadtner_investigating_2023, mejtoft_design_2019}, in practice, designers aim for effective user journeys that meet their goals. Online shopping sites, for example, take advantage of the aforementioned principles~\cite{mathur_dark_2019, chen_practitioners_2022}; they steer users toward effortlessly discovering (recommended) products from where they are quickly manoeuvred to a checkout page~\cite{mathur_dark_2019}. In other cases, design friction is indeed used to hinder certain interactions. This can be sensible to support reflection of an interaction's consequences by interrupting mindless engagement~\cite{cox_friction_2016, mejtoft_design_2019}. However, design friction can lead to frustration in other cases. Along similar lines, infamous cookie-consent banners have a history of making selection processes a cumbersome task~\cite{gray_dark_2021}. They rely on visual interferences and aesthetic manipulation to pre-select choices, effectively pushing users' decisions to their disadvantage~\cite{gray_dark_2021}. 

% responsibility when designing by exploiting psychological errors; dark patternrs
%In these real-world examples, framing effect and the default effect~\cite{}, have been used to misguide and trick users in both e-commerce and cookie-consent banners.  
%In recent efforts to landscape online interfaces for unethical, deceptive design patterns, deploy dark patterns~\cite{} that obscure choices or hinder relevant information from being taken into account. Within the scope of this discourse, it has been suggested that the exploitation of cognitive biases lies at the root of such unethical designs~\cite{}, whether deployed intentionally or not.
% gap: exploring the relationship between cognitive biases and dark patterns
These examples demonstrate the responsibility contained in designers' work. Exploitation of cognitive biases, such as the framing effect or the default effect, have been used to misguide and trick users~\cite{mathur_dark_2019}.
Drawing from the importance of this ongoing discourse and the need for a better understanding of the underlying cognitive mechanisms that can be exploited to harm users, this research aims to take the next step by exploring the relationship between cognitive biases and dark patterns and how it should be navigated in interaction design. To this end, we conducted four focus groups with experts in dark pattern scholarship, cognitive science, and psychology research, several of whom also have significant experience in interaction design ($N=15$). Each focus group included a discussion structured to investigate the following research question:

\begin{enumerate}
    \item[\textbf{RQ:}] How can we conceptualise the dynamic relationship between cognitive biases and deceptive design patterns?
\end{enumerate}
    
%Conclusively, the focus groups helped to connect important insights from two different fields, shaping future work and ethical design considerations. By considering the intrinsic nature of cognitive biases~\cite{tversky_judgement_1974} and the deployment of deceptive design patterns in digital environments, we gained a new understanding of the emerging threats of deceptive design patterns and offer novel insights into underlying mechanisms that designers and policy-makers can build upon to create safer, ethical interfaces. Aiding future work in this field, we contribute the ``Relationship Model of Cognitive Biases and Deceptive Design Patterns'' to describe how design decisions lead to real-world implications and, hence, require responsible decisions to avoid harming end-users.

In conclusion, the focus groups facilitated valuable insights by bridging the perspectives of two distinct fields, providing important details to understand the effects of cognitive biases on dark patterns and vice versa. Moreover, our findings offer guidance for future work and ethical design considerations to better protect users. By examining the inherent characteristics of cognitive biases~\cite{tversky_judgement_1974} and their utilisation in dark patterns within digital contexts, we have garnered new insights into underlying mechanisms, thereby supporting previous suggestions that linked the harming effects of dark patterns to the exploitation of cognitive biases~\cite{mathur_dark_2019,waldman_cognitive_2020}.
% ultimately enabling and leading to deceptive design patterns. %, offering considerations and implications for ethical interfaces. 
Our contribution to this field is the \textit{Relationship Model of Cognitive Biases and Dark Patterns}, which elucidates the connections between design decisions and real-world consequences, emphasizing the necessity of responsible design choices and decision-making to protect end-users.

% contribution:
%The way we think is hugely impacted by cognitive biases and heuristics, that constantly affect our decision making. As they are natural parts of our cognition, they are not inherently bad or good. Historically, biases and heuristics help us to make efficient decisions instead of overthinking minimal problems. However, it research in psychology established a well-understood and accepted list of cognitive biases and heuristics that often lead to sub-optimal decisions and even harming ones. Particular risks arise if these core parts of or inner psych is externally exploited to enforce poor decision making in order to gain benefits. While research of psychologists and linguists demonstrate how unethical and immoral intents easily manipulate or thinking through exploitation of such factors, the consideration of unethical design is relatively new in the field of HCI. Yet, through the concept of dark patterns, the past decade has brought forward a growing taxonomy of malicious interface tricks that alter users' decision-making.

%%%%%%%%%%%%%%%%%%%%%%%%%%%%%
%       RELATED WORK        %
%%%%%%%%%%%%%%%%%%%%%%%%%%%%%
\section{Related Work}
This paper synthesises research on autonomy and empowerment, cognitive biases, and dark patterns. First, we revisit contributions about users' autonomy within the periphery of HCI promoting ethical design concepts. Second, we present a background of cognitive bias scholarship that informed this research. Concluding this section, the third part focuses on HCI research on dark patterns and points to an ongoing discourse regarding terminology in this area.

% First: User Autonomy
\subsection{User Autonomy and Empowerment}
% Value-sensitive design
As digital technologies have become increasingly ubiquitous, the relationship people share with their daily drivers has been described as complex and emotional~\cite{terzimehic_tale_2023}.
The field of HCI has advanced from focusing on usability issues to designing technologies that foster positive interactions and user well-being, often navigating persuasion and autonomy. As an antagonist to autonomy, persuasive technologies often undermine user agency~\cite{chamorro_ethical_2023}. While research in health-related environments (e.g.~\cite{alexandrovsky_serious_2021}) indicates potential positive applications of the persuasive technology paradigm, it is crucial to remain mindful of the ongoing critiques and ethical considerations surrounding persuasive technology. At the same time, it is important to recognise that participants made an autonomous decision and consented before giving up autonomy to change their behaviour toward more healthy options using this health-care intervention. 

Unfortunately, achieving responsible design is not an easy task~\cite{grimpe_towards_2014}. While Schneider et al.~\cite{schneider_empowerment_2018} emphasised the importance of empowerment based on a structured literature review, persuasive design remains an often relied-on strategy for behaviour change interventions~\cite{hutchison_persuasive_2014}. Thereby, the implementation of persuasive design is often well–meant but executed in a problematic manner~\cite{chamorro_ethical_2023}. To illustrate, in their work, Brynjarsdóttir et al.~\cite{brynjarsdottir_sustainably_2012} describe the pitfalls of persuasive sustainability --- the attempt to persuade users' behaviour toward environmental sustainability, related to Fogg's Behaviour Model~\cite{fogg_persuasive_2009}. In sum, Brynjarsdóttir et al. frame a critic concerned with modernist technologies, which they place persuasive sustainability under, and remind about a lack of awareness of a design's impact. Opening future avenues, the authors suggest that, firstly, persuasion and users' reactions have to be better understood before implementation. Secondly, they echo traditional user-centred design (UCD) approaches by reminding practitioners to involve users throughout development stages, and, thirdly, encourage practitioners to move away from the individual and consider larger social environments.
Offering both research and industry a reflection tool in this vein, Elsayed-Ali et al.~\cite{elsayed-ali_responsible_2023} created an online card tool showcasing a variety of critical aspects and considerations for innovative design processes. To that end, the tool fosters the sharing of opinions across hierarchies, accounts for differences among participants, promotes inclusiveness, and gives room to otherwise difficult discussions specific to the participants' work environments. 
 
Work in HCI has long been concerned with conscious control~\cite{jain_designing_2020}, and in that sense user autonomy and agency; although often using these terms interchangeably, as highlighted by a recent literature review~\cite{bennett_how_2023} considering over 32 years of related work. In this review, Bennet et al.~\cite{bennett_how_2023} allocate these values as key concepts in various HCI work. 
% Responsible Research and Innovation
In 2014, Grimpe et al.~\cite{grimpe_towards_2014} highlighted the difficulties of responsible design, problematizing the challenge into four core issues: \textit{reflexivity}; \textit{responsibility and responsiveness}; \textit{inclusion}; and \textit{anticipation}. A core effort promoting user autonomy when interacting with technologies is value sensitive design (VSD), first proposed by Friedman et al.~\cite{friedman_value_2013}. Continuing former work~\cite{friedman_software_1997}, the authors encourage interfaces to convey functionalities transparently and truthfully while empowering users. The relevance of VSD is again highlighted by contemporary work from Chen et al.~\cite{chen_practitioners_2022}. The authors apply underlying paradigms to often deceptive recommender systems to study notions of disagreement between practitioners and users, illuminating disagreement of associated values. In part, these disagreements may be the result of conflicting ethical stances among practitioners and corporate incentives~\cite{gray_ethical_2019,chivukula_wrangling_2023}.

Resulting interfaces might restrict users' ability to make informed decisions that are in line with their beliefs or values~\cite{wang_i_2011,wang_privacy_2013,fogg_persuasive_2009}. %Moreover, studies have found deceptive design to be rather difficult to avoid by end-users~\cite{di_geronimo_ui_2020,bongard-blanchy_i_2021,mildner_about_2023}.
In order to return agency to the user, design interventions, design friction, or nudges have been discussed as potential counter-measures to mitigate loss-of-control feelings getting users to reflect on their choices~\cite{leimstadtner_investigating_2023,hansen_nudge_2013,wang_privacy_2013}. 
Focusing on the social medium Facebook, Lyngs et al.~\cite{lyngs_2020} compared two design interventions to empower users to reflect on their usage behaviour. One would periodically remind the users about their initial goals; the other removed the newsfeed feature from Facebook to keep users' on track with their goals. In a controlled setup, the study revealed certain shortcomings but highlighted how interventions can help users reflect on their behaviour. Exploring how nudges can be used similarly, Masaki et al. \cite{masaki_exploring_2020} developed nudging interventions deployed to shield users' privacy. These findings gain support by work conducted by Wang et al.~\cite{wang_effects_2014}, demonstrating the positive effects of design interventions, friction, and nudges that can grant users better agency when engaging with social media. In a similar vein, Lukoff et al.~\cite{lukoff_how_2021} conducted a co-design study based on YouTube's mobile interface to develop alternative mechanisms that redirect a sense of agency to the platform's users. The limited agency of social networking service (SNS) users could be an indicator of decreased well-being. While previous studies promote design interventions as counter-measures to problematic interfaces in this regard, we return to traditional UCD principles aiming to understand users' expectations regarding UI features within and beyond the SNS context. Exploring the possibilities for adaptive interfaces users can customise to meet their needs, Kollnig et al.~\cite{kollnig_we_2023} developed an ``app repair framework'', after a majority of 85\% of their participants demonstrated appreciation for the option to alter elements of their apps. With a focus on audio-related privacy when recording, Dunbar et al.~\cite{dunbar_listening_2021} propose design principles to accommodate people's concerns and preferences in this sensitive context.
These examples highlight the necessity of understanding when and how in the design process to factor in considerations regarding user agency.

% Our extension
Inspired by this body of work, our objective is to extend our focus beyond specific usage contexts and conceptualise the nuances of design choices practitioners encounter. Furthermore, we intend to incorporate ethical considerations from dark pattern scholarship into the discourse at a more general level. To that end, we connect the fields of cognitive biases and dark patterns to explore their relationships that often lead to negative and harmful effects for users.

% Cognitive Biases
\subsection{Cognitive Biases}
%Design impact in general.
%By nature, design impacts peoples' mental model of an object, where a great design narrates features and characteristics efficiently (and non-verbally) to aid effective engagement. 
%Originally coined by \citet{gibson_systems_1966}, `affordances' describe design elements that (silently) communicate possible interactions. Carried on by Norman~\cite{norman_cognitive_1986}, the concept became a standard within HCI together with `signifiers' that assist this purpose by guiding users' perception toward an object's functionalities. Clever implementation of `feedback' then allows recognition of a system's state allowing users to react towards required or necessary interactions. These design concepts, thereby, rely on human perception and cognition.

% Classics
Utilisation of human cognition and perception is an integral part of human factors and HCI~\cite{wickens_introduction_2003}. In this vein, Jain and Horowitz et al.~\cite{jain_designing_2020} discuss opportunities that go beyond interactions requiring users to make conscious decisions. As the work rests its focus on HCI-related work to promote seamless, unconscious interactions that could enhance people's experience with novel technologies, they remind of ethical constraints when designing to alter cognitive processes. While our work is inspired by underlying constructs of cognitive biases, governing our choices unconsciously, the overall history of this field exceeds the means of this paper. Here, we synthesise key contributions as a background of our work. 

Tversky and Kahneman's~\cite{kahneman_perspective_2003} first introduced the term cognitive biases in 1972. Ever since, related work has identified and described a plethora of effects that influence human decision-making. The granularity and diversity of cognitive biases are well reflected in Baron's book `Thinking and Deciding'~\cite{baron_thinking_2008}, highlighting the enormous effort that went into this field of research, reporting that the variety of concepts has been thoroughly catalogued and understood. In an attempt to summarise these past efforts, Hilbert~\cite{hilbert_toward_2012} identified eight core biases, provoking the idea that every other cognitive bias can be mapped to one of them. Hilbert placed these core biases alongside additional mathematical definitions to avoid any future ambiguity. His framework can help to understand the logical constraints in cognitive processes and promotes the concept of ``The Noisy Memory Channel,'' demonstrating how noises --- confusion and mistakes --- affect decision-making.

With the aim to help people make informed decisions, Hertel et al.~\cite{hertel_cognitive_2011} present a comprehensive overview of cognitive bias modification (CBM) strategies, which encompass procedures designed to prohibit automatic cognitive processes. This includes the alternation of peoples' attention, interpretation of situations, and their memory to affect future decisions. Based on the growing body of work investigating CBM, Jones et al.~\cite{jones_cognitive_2017} conducted a systematic literature review. The authors demonstrate the effectiveness of CBM across selected studies, which consistently show that CBM can modify targeted biases in adults but not children. This work underlines the potential of CBM as a tool for influencing automatic decision processes and reducing cognitive biases. Between the lines, these works postulate certain negative connotations and consequences of cognitive biases. Haselton et al.~\cite{haselton_paranoid_2006} spotlight certain situations where the opposite occurs. Their work on Error Management Theory (EMT) asserts that biased judgments in uncertain conditions can actually result in better decisions compared to unbiased alternatives as part of ongoing research studying the evolution of cognitive biases~\cite{haselton_evolution_2015}.

While these different perspectives mirror the power with which cognitive biases steer decisions, work illustrates the ease in which they can be exploited, for instance, to personalise recommender systems~\cite{theocharous_personalizing_2019} for increasing purchasing behaviour or govern peoples' search behaviour~\cite{azzopardi_cognitive_2021}. To explore possibilities to account for cognitive biases to avoid harmful designs, work by Kahneman and Tversky~\cite{tversky_judgement_1974} and Baron~\cite{baron_thinking_2008} informed the design of our focus groups. We relied on the general definitions and selected biases from these works as part of each focus group. We thus base our findings on the fundamentals of cognitive bias and heuristic decision-making theory to learn about their relation to dark patterns.

\subsection{Dark Patterns}
Critique on nudges~\cite{hansen_nudge_2013} spotlights ethical design caveats to obfuscate people's decision-making. In the scope of HCI, these concerns have sparked various streams of research to investigate the negative or harmful effects of technological interactions --- in its midst a discourse surrounding dark patterns. Introduced by Brignull in 2010~\cite{brignull2015dark}, the concept describes unethical practices that trick users into undesired actions with negative or harmful consequences.
Alongside this discourse, research in HCI has fostered a growing typology of unethical graphical user interfaces fitting the definition of dark patterns across digital media. These include, but are not limited to, games~\cite{zagal_dark_2013} and social media~\cite{mildner_about_2023, schaffner_understanding_2022}, as well as context-specific instances where language barriers are exploited~\cite{hidaka_linguistic_2023}. This effort has recently led to the development of a first ontology of related design patterns~\cite{gray_ontology_2024}. \textit{Interface Interference}~\cite{gray_dark_2018}, for instance, promotes certain interface elements, often visually, to gain users' attention with instances identified in e-commerce~\cite{mathur_dark_2019} and social media~\cite{mildner_defending_2023}. Collectively, these works demonstrate the multitude of deceptive and manipulative strategies practitioners use to steer users' decisions against their best interest and toward service providers' goals. To offer an overview, Mathur et al.~\cite{mathur_what_2021} constructed overarching characteristics by reviewing over 82 types of strategies that include research and regulation-based types. Mildner et al.~\cite{mildner_about_2023} studied the effects of 80 empirically studied types in the context of social media, while Gray et al.~\cite{gray_ontology_2024} considered a corpus of over 245 for their ontology.

While studies have frequently proposed a relationship between dark patterns and cognitive biases~\cite{mathur_dark_2019,waldman_cognitive_2020}, to our knowledge, we are among the first to explore this relationship in a dedicated study. Allowing interpretation in this vein, various studies convey difficulties among their participants to effectively recognise and avoid the effects of dark patterns~\cite{maier_dark_2020, bongard-blanchy_i_2021, di_geronimo_ui_2020, mildner_defending_2023}. Aiming to create privacy-protecting interventions, work has deployed design strategies as so-called ``bright patterns''~\cite{grasl_dark_2021} that invert the mechanisms of dark patterns. The study demonstrates an arguably positive impact on users' decisions. 
However, the study also shows that users' choice architecture is similarly altered as it would be by dark patterns and, thus, restricting users' autonomy to make informed decisions. This aligns with research conducted by Ahuja and Kumar ~\cite{ahuja_conceptualizations_2022}, who demonstrate how dark patterns restrict user autonomy on various levels, bridging the important topics. As the overall goal of this paper is to understand the underlying mechanisms of autonomy-respecting design, we build on their efforts and focus on cognitive and behavioural aspects, particularly nudging and cognitive biases.
% Our extension

\subsection{Terminology}
Recent voices within the dark pattern community have argued that contemporary terminology --- ``dark pattern'' --- should not be used as it poses potential racial misconception~\cite{acm_words_2023}. In an attempt to offer an alternative, Brignull~\cite{brignull2015dark}, who originally coined the concept, suggested using ``deceptive design'' to describe the unethical design practices. However, voices objected to this change~\cite{obi_lets_2022}, arguing the term ``dark'' does not insinuate a ``bad'' interaction but suggests hidden consequences. Furthermore, it dismisses a connection to pattern language~\cite{alexander_pattern_1977}, which conceptualises reusable design strategies and offers solutions to similar problems. This has led researchers to come up with different terms~\cite{monge_roffarello_defining_2023} to describe the same concept. 
While this discourse still unfolds, we opt to use the term ``dark patterns'', staying coherent with related work placing its origin in \textit{hidden consequences} of interfaces instead of evilness or mal-intent, but acknowledge the importance of the current discussion problematising the term. %that it does not perfectly reflect manipulative, coercive, or obfuscating strategies that belong to the same concept. %However, participants of the focus groups used the ``dark patterns'' term repeatedly in the context of ``evil'' and ``malicious'' instead of hidden or obscured design. %suggesting the need for an unbiased term.

%%%%%%%%%%%%%%%%%%%%%%%%%%%%%
%          METHOD           %
%%%%%%%%%%%%%%%%%%%%%%%%%%%%%
\section{Method}
To answer our research question and conceptualise the dynamic relationship between cognitive biases and dark patterns, we conducted a focus group study with experts in dark patterns and cognitive science/psychology research. Several participants in the study also bring extensive experience in interaction design, enriching our understanding of the practical implications in this field. We opted for expert participants to accommodate the exploratory nature of our research question. Additionally, the engagement of these two relevant perspectives in focused discussions promises novel and interesting insights into the similarities, differences, and facilitators of the two concepts. For each focus group, we invited two expert participants with extensive knowledge of dark pattern scholarship and another two with academic backgrounds in psychology or cognitive science. After agreeing to participate, we sent relevant study information to each participant. At the same time, we gained their consent to record and analyse each session in line with the host university's guidelines. Except for one focus group featuring three participants, each was attended by four experts, resulting in a total of 15 participants. To accommodate their international backgrounds and the different time zones they were living in, the focus groups were held online via the video conference tool Zoom.

\subsection{Participants}
\input{tables/participants}
Participants were recruited from the authors' professional and academic networks or word of mouth. We carefully selected individuals with backgrounds and expertise in dark patterns and cognitive science or psychology with the precondition of having published in esteemed venues of their respective fields. Participation was entirely voluntary and without compensation. Before participating in the focus groups, participants were sufficiently informed about the study's purpose and design, about their rights following GDPR guidelines and then asked to give their informed consent. In total, we recruited 15 participants, six self-identified as female, eight as male, and one as non-binary. The average years of experience participants had in their fields at the time of conducting this study was 6.73 years ($sd=3.43$). On average, participants were 32.87 years old ($SD=6.08$). At the time of running this experiment, their professions included (Assistant) Professors (5), Postdocs/Senior Researchers (5), and PhD candidates (5). The demographics of our participants are summarised in Table~\ref{tab:participants}. %Table~\ref{tab:participants} summarises a full overview of our participants' demographics.

\subsection{Focus Groups}
\begin{figure*}[t!]
    \centering
    \includegraphics[width=\textwidth]{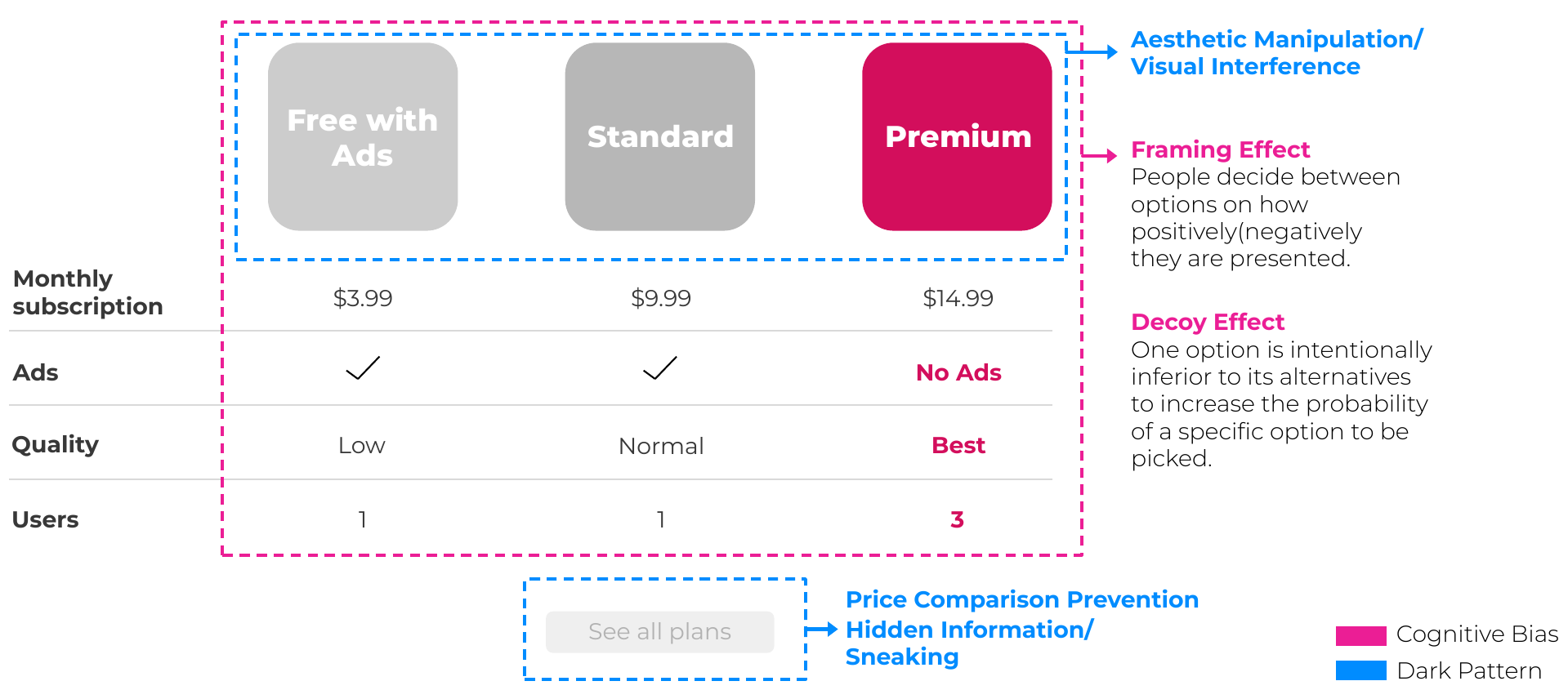}
    \caption{This Figure was part of the introduction in each focus group and illustrates a common monthly subscription model of service providers with different tiers. The figure highlights in magenta the presence of cognitive biases (framing effect and decoy effect) and in blue dark patterns (aesthetic manipulation/visual interference, price comparison prevention, and hidden information/sneaking).}
    \Description{
    Figure 1: This Figure illustrates a three tier subscription model of online services and information for users to compare each tier. However, visual and textual cues steer users' awareness toward a premium option while other tiers are obfuscated through lower contrast colours and/or lighter font sizes.
    }
    \label{fig:example}
\end{figure*}

In the course of 90 minutes, each focus group followed the same procedure: After a brief introduction of all participants, commonly used definitions of the terms \textit{cognitive bias} and \textit{dark pattern} were provided in the form of an online presentation via Zoom. To illustrate previously discussed connections between the concepts, we showed an example illustrating common streaming subscription tiers, including the `framing effect' and `decoy effect' to demonstrate their enabling of \textit{aesthetic manipulation}, \textit{visual interference} as well as \textit{price comparison prevention}, \textit{hidden information}, and \textit{sneaking} dark patterns. Figure~\ref{fig:example} shows the example given during this introduction.

\paragraph{Initial Discussion}
Afterwards, the group was asked three questions to be discussed within ten minutes each: (1) What are the key similarities between cognitive biases and dark patterns? (2) What are the key differences between cognitive biases and dark patterns? (3) What does a cognitive bias facilitate to become a dark pattern, especially with regard to design? Each question affords participants to consider the topic from a different angle. The participants were prompted to use the communication channel of their choice to voice their opinions. They could make use of the chat function of the video conferencing platform or voice their perspective directly via the audio channel.

\paragraph{Card Sorting Task}
After the initial discussion, participants were coupled with a disciplinary counterpart to conduct a card sorting task. Using the online collaboration tool Miro, both pairings~\footnote{For the group featuring three participants instead of four, we joined all experts within a single group to complete the card sorting task together.} were then tasked to group selected cognitive biases and dark patterns within 20 minutes.
To afford the timely constraints of each focus group, we chose to select only prominent cognitive biases and dark patterns based on the citation counts of respective publications. Because of this limitation, however, we did not analyse these results further to avoid a potentially strong selection bias. To give participants more context, cards included definitions taken from original publications.
Despite its limitations, we included this exercise as we wanted to engage participants in a transdisciplinary activity to experiment and test previously discussed ideas. Further, our aim was to inspire alternative perspectives and enrich the following discussion. The supplementary material contains a snapshot of the initial Miro board, including the selected cognitive biases (sourced from Tversky's work~\cite{tversky_judgement_1974}) and dark pattern types (sourced from the dark pattern ontology by Gray et al.~\cite{gray_ontology_2024}).

\paragraph{Reflective Discussion}
Following up on the card sorting task, participants discussed their experiences, difficulties, and ideas behind creating groups. In addition, the experience gained in the previous activity led to new insights with regard to the questions raised in the initial discussion. This allowed a more critical reflection on earlier statements as well as identifying links between cognitive biases and dark patterns that had gone unnoticed before.

%%%%%%%%%%%%%%%%%%%%%%%%%%%%%
%         ANALYSIS          %
%%%%%%%%%%%%%%%%%%%%%%%%%%%%%
\section{Analysis}
Once we completed the four focus groups, we manually transcribed the discussions and prepared the data for further analysis. To this end, we anonymised any content that could be traced back to any individual participant. For the analysis of the transcribed data, we conducted a thematic analysis and used the card sorting results as an assistive source for a better understanding of participants' perspectives. These steps were concluded using the software Atlas.ti~\cite{atlasti_2021}.
%\subsection{How we generated the model}
% based on the focus groups
% different cohorts voiced similar concerns
% hierarchische analyse
%Prior to the analysis, we transcribed and anonymised the focus group discussions.% Using the qualitative data analysis software Atlas.ti, we followed an inductive approach. After iterative discussions, we constructed an initial codebook featuring 52 codes. The codebook can be found in the supplementary material. The 52 codes were then further analysed through axial coding and discussed among the authors, distilling connections between cognitive biases and the emergence of harmful design. 
\subsection{Positionality}
The authors of this research have mixed backgrounds. One author gained their education in West Africa while the others acquired their education in Central Europe with WEIRD (Western, Educated, Industrialised, Rich, and Democratic)~\cite{henrich_weird_2010} backgrounds. Their research expertise includes design, computer science, and psychology, while their academic work focuses on topics concerned with social justice and user well-being in digital technology contexts.
%The authors of this research have mixed backgrounds. Some acquired their education in Central Europe with WEIRD (Western, Educated, Industrialised, Rich, and Democratic)~\cite{henrich_weird_2010} backgrounds, One author gained their education in West Africa. Their research expertise includes design, computer science, and psychology. while their academic work focuses on topics concerned with social justice and user well-being in digital technology contexts.
The focus groups were transcribed by two authors and, afterwards, coded by two authors. All participants were recruited through professional networks and by word-of-mouth. Each focus group was conducted by at least two authors, one of which would be responsible for moderating, the other(s) held assistive role(s). As this research aims to describe the relationship between cognitive biases and dark patterns, this lens has guided the structure and analysis of the focus groups. Finally, we acknowledge any possible biases that are the result of our academic, cultural, and personal backgrounds.

\subsection{Coding of the Focus Group Transcripts}
%Transcripts were manually done... yada yada. Some data about length of each focus group
Prior to the analysis, we manually transcribed and anonymised the focus group discussions. %Using the qualitative data analysis software Atlas.ti, we followed an inductive approach to assess the transcriptions. After iterative discussions, we constructed an initial codebook featuring 52 codes. The codebook can be found in the supplementary material. The 52 codes were then further analysed through axial coding and discussed among the authors, distilling connections between cognitive biases and the emergence of harmful design. 
In the first step, two authors coded a representative sample of 50\% of the material using open coding in line with Blandford et al.~\cite{blandford_qualitative_2016}. We then conducted an iterative discussion to establish an initial coding tree. The remaining transcripts were split between the two authors and
coded individually. Finally, we conducted a concluding discussion
session to finalise the coding tree. This was followed by a thematic analysis to identify emerging dimensions from the material as described by Blandford et al.~\cite{blandford_qualitative_2016}. The codebook used to analyse the transcripts of the focus groups comprises 52 codes and is included in this paper's supplementary material.

%Describe use of the card-sorting task to reflect on prior discussions and give further insights into relationship angles.
The main purpose of the card sorting task was to engage participants in an interactive task where dark pattern scholars and psychologists/cognitive scientists worked together to engage with and analyse their ideas from previous discussions. Hence, the set of biases and dark patterns used for the card sorting task was limited and only included subsets of the overall cognitive bias and dark pattern typologies. Thus, results were used to build an understanding of the data collected in the different discussion phases, therefore aiding our analysis of the focus group transcripts and the following identification of relationships.

%\subsection{Structuring of Codes}
Using axial coding, two authors began to independently organise all codes systematically into hierarchical structures and groups~\cite{creswell_research_2009,blandford_qualitative_2016}. This process was extended by connecting individual codes to describe their relationships, noting whether any confirming or contradicting notions exist between them. At this stage, the authors frequently revisited transcripts and card sorting results to ensure they stayed truthful to the data. Afterwards, the authors exchanged their findings in a following discussion. Here, the emergent hierarchical frameworks of the codes were the focus of identifying overarching themes. We thereby aimed to identify phases that followed design practice to real-world consequences in terms of dark patterns. Consequently, based on the analysis outlined above, we constructed a model that encapsulates these phases, effectively bridging the gap between design practices and their tangible impacts in the context of dark patterns.

%%%%%%%%%%%%%%%%%%%%%%%%%%%%%
%         FINDINGS          %
%%%%%%%%%%%%%%%%%%%%%%%%%%%%%
\section{Findings}
Based on the thematic analysis, we gained certain insights into the relationship between cognitive biases and dark patterns. However, this relationship appears intricate and, expanding prior suggestions~\cite{mathur_dark_2019,waldman_cognitive_2020}, multifaceted. Here, we echo our participants' discussions to provide common and specific characteristics of the two fields. Moreover, we focus on the implications of design, the inscription of functionalities, and real-world applications as discussed by our participants, and promote the ``Relationship Model of Cognitive Biases and Dark Patterns''. Our presentation of the results as a model is consistent with established practices of presenting results in HCI and Ubicomp (e.g. %the lived informatics model of personal informatics,
~\cite{epstein_lived_2015,jain_designing_2020,dunbar_listening_2021,mildner_listening_2024}).

\subsection{Similarities, Differences, and Facilitators}
Each focus group included granulated discussions about similarities, differences, and facilitation between cognitive biases and dark patterns. Although the participants either had backgrounds in dark pattern scholarship or cognitive science/psychology research, combined with experience in interaction design, providing definitions and a visual example established a common ground for fruitful discourse.

\subsubsection{Similarities}
When asked about the similarities between the two topics, participants noticed shared attributes regarding the impact on decision-making and autonomy. Arguing from a designer's perspective, P5 discussed this sentiment further when persuading user actions:
\begin{quote}
    \textit{``I think the key similarity is [...] how the steering happens. [...] You are trying to steer your users from the normative ways of user agency, which is what HCI design mostly preaches.'' -- P14}
\end{quote}
By leveraging users' perception and cognition, responsibility plays a pivotal role in the profession of designers. In this regard, P3 noticed another similarity in the shared dangers and harms that occur when not cared for. The same participant went on to discuss the particularities of users' unawareness:
\begin{quote}
    \textit{``[B]oth can be unconscious. So, in both cases, the user may not be aware of the influence on their decision-making.'' -- P3}
\end{quote}
Overall, participants noticed certain synergies between cognitive biases and dark patterns. When triggered, both carry risks of making unfavourable choices as either is difficult to avoid. Certain responsibility was further attributed to designers utilising cognitive biases when used to steer users against their will.

\subsubsection{Differences}
The second question sought to identify differences between cognitive biases and dark patterns. Across all focus groups, a dominant argument addressed the different natures of either concept. While  \textit{``biases are already there [...] and inform our decisions''}, as P9 pointed out, or \textit{``your intuitive brain [...] working in free-flow without really engaging too much in reasoned thought''} (P1), dark patterns, on the other side, are actively created and deployed by practitioners. P1 later continued their argument in consideration of the power dynamic between the designer and user:
\begin{quote}
    \textit{``[A] designer has complete control over how that pattern takes form, takes shape, and how it's implemented within a system. Whereas a user doesn't have control over their cognitive biases.'' -- P1}
\end{quote}
As cognitive biases are intrinsic to our behaviour, dark patterns, like all design patterns, are created and impact our behaviour intrinsically. Importantly, participants noticed that not all dark patterns actually require cognitive biases. This is in line with a statement made by P10:
\begin{quote}
    \textit{``[O]bfuscation and sneak-into-basket and hidden costs, all of those just seem like outright lying. So, it's not necessarily using a cognitive bias to be the problem.'' -- P10}
\end{quote}

\subsubsection{Facilitators}
During the last question, participants explored possible facilitators between the concepts. As similarities and differences foreshadow related attributes, P9 discusses how dark patterns emerge:
\begin{quote}
    \textit{``[D]ark patterns are really the design material that allows those cognitive biases to be activated [...] this method is used to take advantage of our cognitive biases.'' -- P9}
\end{quote}
A strong sentiment across participants for a facilitator was further noticed in (mal)intent behind deploying dark patterns. In this regard, P3, P12, and P15, thoroughly discussed designers' roles in persuading users' decisions toward commercial goals, limiting their autonomy. Sharing this position, P10 noted:  
\begin{quote}
    \textit{``Something is dark patterns when you use any means whatsoever [to] deprive me of my autonomy or to [...] engage me in practices that violate my privacy.'' -- P9}
\end{quote}
Although not all dark patterns rely on cognitive biases, the latter seems to be an effective means to the end for malicious strategies. This reconnects to the previously discussed responsibilities of designers but also highlights additional needs for ensuring user safety. As P3 anecdotally pointed out, however, regulations, such as the GDPR, do not consider intent as a necessity. This would ensure user protection whenever harm is done.

%%%%%%%%%%%%%%%%%%%%%%%%%%%%%%%%%%%%%%%%%%%%%%%%%%%%%%%%%%%%%%%%%%%%%
%  The Relationship Model of Cognitive Biases and Dark Patterns.    %
%%%%%%%%%%%%%%%%%%%%%%%%%%%%%%%%%%%%%%%%%%%%%%%%%%%%%%%%%%%%%%%%%%%%%
\subsection{The Relationship Model of Cognitive Biases and Dark Patterns}
To address the echoed implications for responsibility and impact for practitioners, we opted for developing a model that follows design from its creation to real-world implications. In a preliminary version, we modelled how cognitive biases and their exploitation can lead to deceptive design. Our model was constructed based on the findings of our study juxtaposed with relevant previous work. We verified our model by inviting focus group participants to provide feedback on individual and general levels of this preliminary model, helping us iterate and improve it where necessary. 
To this end, we sent out an online survey to all participants, informing them about the task and gaining their consent before breaking down the model for individual critique. Additionally, we offered participants to reach out to us and share further comments. In total, four participants responded to the survey, helping us to advance the model. %RWe analysed their accumulated feedback and made changes to the preliminary model accordingly.

\begin{figure*}[t!]
    \centering
    \includegraphics[width=\textwidth]{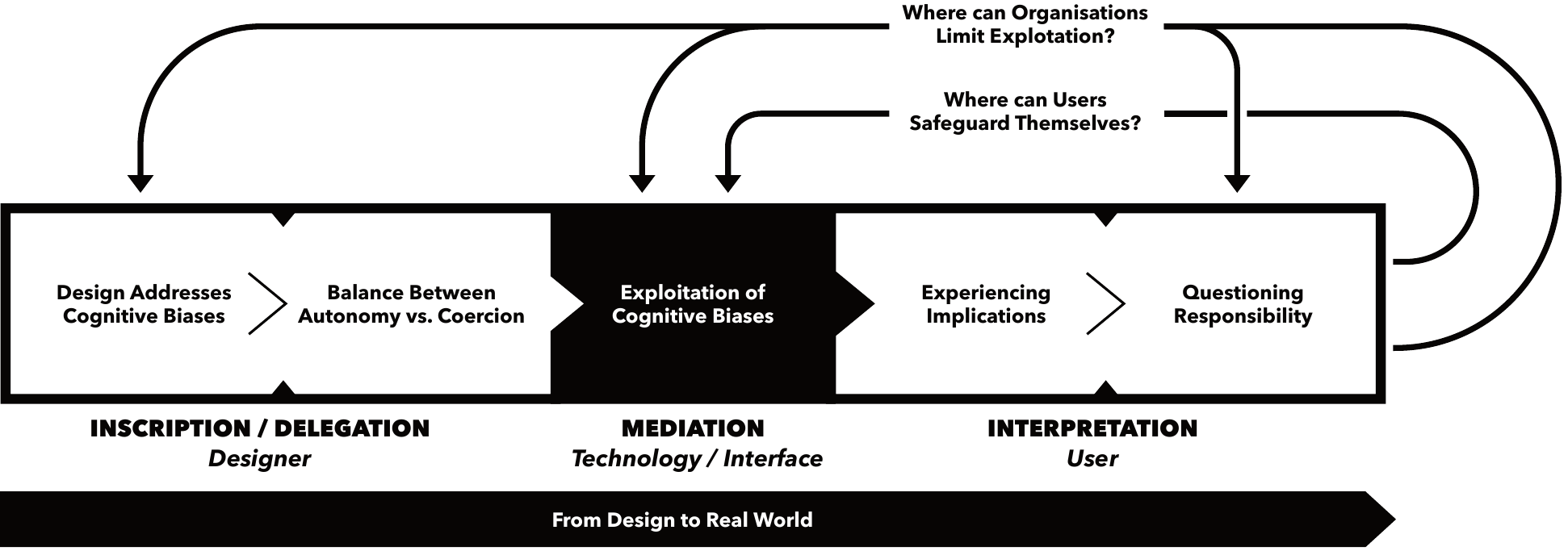}
    \caption{This Figure presents the \textit{Relationship Model of Cognitive Biases and Dark Patterns}. Following a continuum from (potentially unethical) design to real-world applications, the model comprises three stages spanning five phases. Adopting Verbeek's theory of technology mediation~\cite{verbeek_what_2005,verbeek_materializing_2006}, the model follows designers' inscription of functionalities into technology to users' interpretation, leading to the questioning of responsibilities. Depending on the impact and implications of the (unethical) design, end-users may need safeguarding measures, while policy and regulation may be required for their protection.}
    \Description{
    Figure 2: This Figure shows the Relationship Model of Cognitive Biases and Dark Patterns. The model follows a continuum from (potentially unethical) design to real-world applications comprising two stages, adopting Verbeek's theory of technology mediation, and five phases, from utilising cognitive biases to ascribing responsibility: (1) utilising cognitive biases; (2) balancing between autonomy versus deception; (3) exploiting cognitive biases; (4) assessing implications; and (5) ascribing responsibility. Depending on the impact and implications of dark patterns, end-users may need safeguarding measures, while policy and regulation may be required for their protection.
    }
    \label{fig:model}
\end{figure*}

Based on the focus groups, collected feedback, and prior work, we introduce the \textit{Relationship Model of Cognitive Biases and Dark Patterns} as an answer to our research question (visualised in Figure~\ref{fig:model}): How can we conceptualise the dynamic relationship between cognitive biases and dark patterns? 
This model comprises three stages that encompass five phases, showcasing a process from design to the real world. More precisely, inspired by ethical considerations of dark patterns~\cite{gray_dark_2021}
%~\footnote{Throughout the focus groups, participants often used the term ``dark pattern'' to describe the deceptive and manipulative practice of steering user decisions. To be truthful to our participants' statements, we decided not to alter the term to ``deceptive design patterns'', as used in the rest of this paper, but used deceptive design patterns throughout the description of our model interchangeably.} 
and their relationship toward cognitive biases~\cite{mathur_dark_2019,waldman_cognitive_2020}, the model follows the implementation of dark patterns (or other deceptive designs) from addressing cognitive biases in design to potential real-world implications, thereby showcasing potential harmful consequences which require the assessment of responsibilities.

Drawing inspiration from Verbeek's theory of technology mediation~\cite{verbeek_what_2005,verbeek_materializing_2006}, our model comprises three stages: (1) Inscription / Delegation from a designer's perspective, (2) mediation of the technology or particular interface, and (3) users' interpretation thereof. Adopting Verbeek's theory, the model demonstrates how dark patterns emerge from exploiting cognitive biases. To this end, our model breaks the three stages down into five phases. 
The first stage describes the designer's perspective to inscribe or delegate functionalities. Happening in phases one and two, \textbf{design addresses particular cognitive biases} determining the \textbf{balance between autonomy versus coercion}. The second stage -- mediation -- contains the third phase: the \textbf{exploitation of cognitive biases}, which can ultimately lead to deceptive practices and harm. The third stage focuses on the users' point of view, interpreting the design throughout the fourth and fifth phases. First, users \textbf{experience the designs' implications}, leading them to \textbf{questioning responsibility}. 
% The five phases of the model are (1) Utilising cognitive biases; (2) balancing between autonomy versus deception; (3) exploiting cognitive biases; (4) assessing implications; and (5) ascribing responsibility. 
Additionally, we identified crossroads for safeguarding strategies of end-users as well as opportunities for organisations to limit harmful design through exploiting cognitive biases. In the following subsections, we outline each phase in detail. We support the descriptions for each phase through quotes from our participants and connect individual phases to related work where applicable. For improved readability, we slightly altered some statements, ensuring words and sentiment were maintained.

\subsection{From Design to Real World}
The aim of the \textit{Relationship Model of Cognitive Biases and Dark Patterns} is to follow the development of (unethical) design from its planning stages until its deployment into the real world. Importantly, as with many things, breaking decisions and their consequences into stages or phases can result in over-simplifying processes. Our model is no exception. To emphasise the responsibility tied to designers' decisions and connect them with the consequences that arise once a design is implemented in the real world, we represented this progression with a continuous arrow beneath the three stages. Research~\cite{gray_dark_2018,mathur_what_2021} and regulation and legislation~\cite{eu_dsa_2022,edpb_guidelines_2022,ccpa_2018} suggest that malintent plays a role in the design of online interfaces to deceive or manipulate users. However, many design decisions are not aimed at exploiting users' cognitive biases, and neither are practitioners always able to predict resulting harms. Throughout development, many constraints (e.g. money and time) limit the possibilities in which a design~\cite{gray_ethical_2019,verbeek_what_2005} can be tested. Over-simplifying these issues can deflect from the critical aspects when design turns deceptive. 
Echoing the voices of our participants, the following phases attempt to convey these critical aspects with the goal of respecting the underlying continuum.

% PHASE 1
\subsection{Phase 1: Design Addresses Cognitive Biases}
The first phase of this model addresses the utilisation of cognitive biases to control users' attention or afford specific interactions. This is in line with Cross' description of design cognition~\cite{cross_design_2001}, where it is part of the design process to identify the right problem to be solved: Depending on the situation and the underlying goal, a design may draw the user toward or away from it. Interactions are either better supported or actively obfuscated to guide users' decisions. While there are plenty of reasons for either strategy, dark patterns are used to provoke certain choices.  It is important to emphasise here that our data show that specific cognitive biases are addressed both consciously (conscious decision of the designer) and unconsciously (not deliberately addressed by the designer).  The \textit{Bad Default} dark pattern~\cite{bosch_tales_2016}, for instance, is often used in privacy settings where the system provider is initially allowed to collect personal information until a user decides otherwise. This pattern finds support in the `default effect' as people often follow existing choices~\cite{baron_thinking_2008}.  Whether the deployment of these strategies and usage of supportive cognitive biases is a deliberate choice is questioned by P14:
%intrinsic
\begin{quote}
    \textit{``Are [practitioners] doing this knowingly? I mean, the obvious is that [if] you try to get people to buy a certain thing, [...] how dark is a dark pattern if it's just accidental?'' -- P14}
\end{quote} 
Designers can address the \textit{intrinsic} nature of cognitive biases --- shared between humans --- when creating or using dark patterns to alter peoples' choice architecture. The particular relationship between intrinsic cognitive biases and extrinsic deployment of dark patterns was repeatedly discussed among participants across the focus groups. P15 and P12 populated this idea through the following statements:
\begin{quote}
    \textit{``[C]ognitive biases are also so embedded in our automatic decision-making process, while dark patterns are our techniques deployed by external entities. [...] One is individual, embedded decision-making, and the other is external to us.'' -- P15}
\end{quote}
\begin{quote}
    \textit{``[C]ognitive biases are the receiver while the [dark pattern] are something from the perpetrator.'' -- P12}
\end{quote}
The extrinsic effectiveness of dark patterns thereby hugely benefits from intrinsic cognitive biases, described by some participants as a symbiotic relationship. The knowledge garnered by cognitive scientists and psychologists plays a large role in informing dark patterns and spotlighting opportune moments to leverage cognitive biases. However, designers are not always in possession of this knowledge. P15 elaborated on the potential of this knowledge by saying:

\begin{quote}
    \textit{``Knowing how we process information and how we heuristically process and think is the way dark patterns become so effective.'' -- P15}
\end{quote}

% PHASE 2
\subsection{Phase 2: Balance Between Autonomy vs. Coercion}
%intrinsic cognitive biases vs extrinsic exploitation in deceptive design
In the second phase, designers navigate the challenging equilibrium between empowering users by giving them autonomy over their choices and deceptive practices that coerce users' decisions.
The (conscious and unconscious) decisions made in the first phase have a direct influence on this balance, where ethical considerations and consequences are introduced. As with the first phase, the balance between providing users with autonomy and coercing their actions can be an active design choice or an unaware consequence~\cite{chivukula_wrangling_2023}. The effect on the user, however, is often the same. Therefore, this phase requires careful attention to support reflective and informed decision-making. P10 argues that it can be dangerous for users when people attribute expertise to themselves in areas they do not actually have and that this can potentially cause a lot of harm, especially in sensitive contexts such as mental health applications:
\begin{quote}
    \textit{``[S]elf assigned expertise [...] gives [practitioners] that right framing that they could then use dark patterns that exploit cognitive biases to design products without feeling like they are being very unethical.''-- P10}
\end{quote}

Instead of designing responsibly with users' best interests in mind, common or best practices and traditional mindsets within professions may foster excuses for harmful design implications at the user's cost~\cite{gray_ethical_2019,chivukula_wrangling_2023}. In this regard, the `status-quo' bias~\cite{baron_thinking_2008} may offer some insight into why practitioners and designers resort to deploying the same problematic designs. As with persuasive design and nudges, however, their intention might be noble, but the outcome is detrimental to users' well-being.
%Moreover, in the case of mal-intent, the malicious or damaging aim is more obvious than in the previously mentioned examples, as work on deceptive design patterns demonstrates. 
The plethora of deployed strategies that harm users~\cite{gray_ontology_2024} is witness to practitioners' negligence in this regard, reflected by the increased regulatory efforts in place to protect users~\cite{eu_dsa_2022,ccpa_2018}. However, it has to be noted that this negligence might stem from a culture of oversight or neglect at the organisational level~\cite{chivukula_wrangling_2023}. This second phase concludes the first stage of our model and the perspective of designers who inscribe their ideas into the design.

% PHASE 3
\subsection{Phase 3: Exploitation of Cognitive Biases}
Relying on practitioners' decisions in the first stage, the second stage contains the mediation of the design and the third phase where exploitation of cognitive biases manifests --- eventually resulting in deceptive practices and harm. It conveys the implications of design that harnesses cognitive biases and controls user behaviour. In other words, this phase is about users' interactions with the design and, thereby, the first phase, where design enters the real world with all its implications. It describes the users' reaction to the design but not yet their evaluation of any interaction.
Dark patterns that govern user interactions influence their choice architecture without providing transparent information about their consequences. In this vein, P9 drew the following connection: 
\begin{quote}
    \textit{``[A] cognitive bias is a necessary component as kind of [...] an ingredient to a dark pattern''. -- P9}
\end{quote}
% essence what happens:

A core element to enable design --- whether for good or bad --- is to afford specific interactions~\cite{norman_design_2013}. To that end, affordances can have many forms and shapes to convey their aims but have to be delivered easily accessible for users to engage. A common choice to create affordances is through nudges~\cite{thaler_nudge_2008}. While work demonstrates their effectiveness~\cite{caraban_23_2019}, they limit transparency as P9 detailed: 

\begin{quote}
    \textit{``[N]udges for good fail to respect the autonomy of the person who's using the system if they're not very open and transparent about how the nudges are working and there becomes a power imbalance then between whoever is providing those nudges and the person on the other end of them''. -- P9}
\end{quote}

This power imbalance is a critical implication of dark patterns as the user is rarely in control or aware of any consequences~\cite{mildner_defending_2023}. Thereby, design can easily facilitate its goals without providing sufficient information that would allow users to reflect on their choices. P10 explained how preferences can be exploited in the form of cognitive biases: 

% design to afford actions;
% nicht so viel bewertung sondern mehr autonomatismen.

\begin{quote}
    \textit{``[U]sing our tendency to prefer colour or be attention-grabbing so that we look at [the interface] more or our tendency to kind of not want to think more than we have to. [...] [D]ark patterns [are] more of a method and cognitive biases [are] more of an inherent thing that is exploited.'' -- P10}
\end{quote}

Our participants also discussed the natural benefits of certain cognitive biases and heuristics to provide effective shortcuts in various situations. In tandem with current design practices and service providers' monetary incentives~\cite{chen_practitioners_2022}, their mere presence opens the door for serious exploitation, limiting users' ability to formulate reflected decisions and avoid manipulation into unfavourable interactions.

\subsection{Phase 4: Experiencing Implications}
The fourth phase of the model addresses the real-world implications behind dark patterns and how users experience them. Here, the third and last stage begins, where users interpret the design. At this time, the impacts of the design become noticeable and perceptible, whether originally accounted for or unintentional. Informed by the first two stages of the model, during the fourth phase, practitioners have already deployed the strategies to deliver their design's goal(s). This includes any decisions that affect user agency and autonomy through cognitive biases. In other words, one could say that from a user's perspective, the harm is done. Elaborating on users' expectations, P10 raises important questions from a user's point of view:

\begin{quote}
    \textit{``[W]hat mental models do I have about these products? And in what ways [do] my mental models of this product align with what's happening before me?'' -- P10}
\end{quote}

Before any interaction, users may have certain expectations of what will happen. The previous phases and underlying choices ultimately lead to the moment when a user interacts with a design for the first time. While certain dark patterns prohibit reflective thinking, others obscure access to specific functionality. Unaware of the practitioners' goal, users rely on a design's attributes to understand required actions. Thereby, they can draw from their mental model and past experience before engaging with it but cannot know for certain what the consequences will be. At this stage, practitioners and users will only be able to assess the truth behind the implications once harm has been done. 

% PHASE 5
\subsection{Phase 5: Questioning Responsibility}
In response to the aftermath of the prior phase, the fifth phase mirrors past decisions and reflects the question of responsibility behind harmful implications, concluding the third stage of interpretation. However, this question is not simply answered. From a user's perspective, technological illiteracy or unawareness of the presence of dark patterns can lead to seeking responsibility in themselves. Practitioners, on the other end, may displace responsibility behind best practices and common, albeit critically viewed, design paradigms or deflect responsibility altogether, convinced of the noble cause of their design. In any case, the responsibility is difficult to ascribe to a single party. Nonetheless, there are multiple avenues to mitigate negative implications. In this regard, P1 reflected on unintended but harmful interactions, taking the practitioners' perspective:
\begin{quote}
    \textit{``[T]here is some kind of control over it from a designer's point of view that once [the harm] has been identified and they've observed the problem, that they can go back and fix it then afterwards''. -- P1}
\end{quote}
However, the strategy outlined by P1 would require a thorough understanding of exploited cognitive biases, alternative strategies, and access to necessary resources to change the design post-development. Especially in this phase, it is important to remember this model also as a continuum where decisions are intertwined and causally dependent. Hence, untangling the root causes behind implications and identifying responsibility is difficult. Therefore, practitioners need to acquire a critical view of their own work and reflect on utilised strategies to understand how cognitive biases enable dark patterns that affect end-users' decision-making. This does not mean that practitioners need to become experts in human cognition and perception (although it would not harm). Reflective steps throughout development stages can already help ensure that user autonomy is sufficiently respected to empower informed decision-making.

\subsection{Safeguarding and Counter Measures}
Collectively, the five phases of the \textit{Relationship Model of Cognitive Biases and Dark Patterns} highlight how dark patterns leverage cognitive biases and how the latter can enable the first. However, counter-measures to existing dark patterns and their harms have not yet been accounted for. We, therefore, expanded the model with two additional arrows that link to opportune moments where users could safeguard themselves and where organisations, such as governmental bodies, could restrict the exploitation of cognitive biases through guidelines and regulations. Suggesting that knowledge can help users mitigate harmful effects, P9 stated the following:

\begin{quote}
    \textit{``Sometimes, if it's raised to your attention, you can fight back yourself. But in other cases, the pull is so strong that you actually need regulatory bodies or other kinds of policies in place to fight back on your behalf.'' --P9}
\end{quote}

Safeguarding measures for and by end-users are only effective if the problems are understood and can be avoided. In this regard, one arrow leaving the fifth phase carries the question: \textit{Where can users safeguard themselves?} --- points to the third phase, where cognitive biases are exploited. Users can learn to mitigate cognitive biases, for instance, through understanding how they manifest~\cite{cockburn_anchoring_2019,hertel_cognitive_2011}. But even then, it may not be possible to always rely on the safeguarding measures as demonstrated by Bongard-Blanchy et al.~\cite{bongard-blanchy_i_2021} or Mildner et al.~\cite{mildner_defending_2023}. Importantly, our participants have emphasised that this is not (ever) the user's fault. Hence, our participants envisioned that knowledgeable people would be able to recognise the potential exploitation of cognitive biases in the third phase and could proceed strategically with an appropriate reaction. P9 later continued their argument by discussing the current approach behind dark patterns and proposing legal requirements to assure user protection:

\begin{quote}
    \textit{``I think a lot of the mainstream dark patterns that seem to be very effective at triggering our cognitive biases and altering our ability to act or understanding what the situation is, do certainly seem like they're illegitimate forms of altering sense-making versus ones that we would describe as normatively acceptable.''--P9}
\end{quote}

In this regard, P3 points toward already existing regulation, such as the  Digital Service Act (DSA)~\cite{eu_dsa_2022} of the European Union (EU), restricting the implementation of dark patterns. Exploitation of cognitive biases, however, is not regulated:

\begin{quote}
    \textit{``There [is] the word dark patterns in the law now, but cognitive biases are not a term for law.'' --P3}
\end{quote}

In our model, organisational limits to exploitation, such as guidelines or regulations, are alternatives to end-user safeguarding. In this regard, the model incorporates a second arrow leaving the fifth phase asking: \textit{Where can organisations limit exploitation?} The arrow links to the first, third, and fifth phases to highlight opportunities to protect users. In the same direction as pointed out by P3, organisations could request user autonomy to be respected, as done by the EU's DSA~\cite{eu_dsa_2022}. They could also restrict the exploitation of cognitive biases through regulating deceptive design, also done by the DSA as well as other regulations such as the US state of California's CCPA~\cite{ccpa_2018} or India's recent issue to prevent dark patterns~\cite{indiaccpa}. Lastly, the arrow loops back into the fifth phase to illustrate the importance of identifying offenders and holding them accountable if users are harmed. 

Although we view these measures as more effective compared to users' safeguarding themselves, we recognise that they are also more drastic and finite in the case organisations increase regulations. On the one hand, they could address the exploitation of certain biases directly, for example, by requiring specific user consent or autonomy to prevent harmful, impulsive interactions. This line of defence leaves space for practitioners to choose alternative design strategies. On the other hand, regulations can restrict the overall available design space, prohibiting the deployment of dark patterns. As with the DSA~\cite{eu_dsa_2022}, specific strategies can be directly addressed to ensure user safety.

%%%%%%%%%%%%%%%%%%%%%%%%%%%%%
%        DISCUSSION         %
%%%%%%%%%%%%%%%%%%%%%%%%%%%%%
\section{Discussion}
Inspired by previous research proposing a connection between cognitive biases and dark patterns~\cite{mathur_dark_2019, waldman_cognitive_2020}, this paper offers answers to our research question by problematising the dynamic and intricate constraints within this relationship. To that end, we conducted four focus groups and, based on our results, proposed the \textit{Relationship Model of Cognitive Biases and Dark Patterns}. This model depicts five phases that describe how decisions to deploy cognitive biases can enable exploitation and dark patterns. In this section, we discuss applications of this model, how technology can be devised to preserve user autonomy in such contexts and point toward future avenues HCI could consider to foster the implementation of ethical design and user protection.

\subsection{Using the Relationship Model of Cognitive Biases and Dark Patterns}
% Practitioner perspective model to support reflection processes in design
Previous work has described a growing typology of dark patterns~\cite{mathur_what_2021, mildner_about_2023, gray_ontology_2024} users constantly encounter across web and app interfaces. While various patterns have been captured and are relatively well understood, this strand of research lacks in-depth knowledge of the underlying mechanisms.
Contributing relevant insights in this regard, the main aim of the \textit{Relationship Model of Cognitive Biases and Dark Patterns} is to support both researchers and practitioners by providing them with reflective phases that can support them in considering the ethical caveats and impacts of their designs. It is not meant to be prescriptive but provides a guiding roadmap reminding about ethical implications throughout the design's lifespan and the interplay of cognitive biases and dark patterns in that regard. 
While the aim of our model cannot change any malicious objectives of practitioners, it can serve as a reminder about the implicated consequences of utilising cognitive biases and can guide toward potential counter-measures.
% - 1st where dark patterns happen
The model can, therefore, be applied in situations where dark patterns are observed. \textbf{Especially in the early development stages of designs, the \textit{Relationship Model of Cognitive Biases and Dark Patterns} can complement decisions made alongside existing, traditional design paradigms that may not always prioritize user agency and autonomy.}

% - prevention of harmful design 
The research community focusing on dark patterns~\cite{gray_dark_2023} has recently questioned the intent behind deploying strategies that harm users. Throughout their discussions, participants across our focus groups separated practitioners' intent from the harmful impacts of deployed interactions. While some practitioners utilise obvious tactics to manipulate users' choices to their service's advantage, smaller service providers may naively follow existing common practices and deploy dark patterns without realising the consequences. However, in both cases, the end-user may suffer from possible implications, regardless of intent. Based on these insights, we argue that \textbf{necessary protective measures should be independent of intentions as they deem the safety of end-users most important. }
% -> policies
Recent regulations~\cite{eu_dsa_2022,ccpa_2018} already try to restrict certain dark patterns. While this offers promising avenues to enhance user protection and safety, we maintain that regulatory measures should be employed judiciously when other protective measures have proven ineffective. The form and shape in which regulation should be enforced pose new, complex challenges. On the one hand, over-regulation may restrict innovation in the early stages. On the other hand, it is difficult to administer regulations that affect domains as large as web and app interactions.
% 1-2 implication statements in bold

\subsection{Preserving User Autonomy}
% using our tool to promote autonomy
A key aspect of our model is to provide arguments for the importance of supporting and preserving user autonomy.% through an empirical analysis. 
Based on our findings, we argue that it is crucial to allow for and foster informed decisions before interactions happen.
% Good intentions bad outcomes
Traditional HCI paradigms offer practitioners a range of tools and common practices to support ethical design practice. Although critically reviewed, persuasive design explains how motivation can be directly addressed to steer and alter users' choice architecture and guide them toward a pre-determined goal. While behaviour change can be supported through design and technology, it is easy to mean good and do harm~\cite{brynjarsdottir_sustainably_2012}.
% - % reflecting back to nudges and persuasion
In research, it is customary to behave responsibly, ask for consent, and make users aware that the technologies utilised will potentially affect their choices. Outside research, users are often kept unaware of the consequences behind interactions, often noticed in infamous cookie-consent banners~\cite{utz_informed_2019, gray_dark_2021}.
% End-user Safeguarding
% - prozessebene: wann user safeguarding -> future work wann safeguarding strategien effektiv sind <- studien zeigen dass es nicht immer geht, aber vielleicht geht es ja doch?
Knowledge about the enabling effects of cognitive biases toward dark patterns could mitigate certain exploitation and protect users~\cite{jones_cognitive_2017}. However, multiple studies~\cite{mildner_defending_2023,bongard-blanchy_i_2021,di_geronimo_ui_2020} investigating the end-users ability to recognise and avoid dark patterns repeatedly demonstrate difficulties among their participants, even if information about dark patterns was provided. This indicates that user safeguarding is limited in its effectiveness, and other means are necessary to uphold their autonomy. 

% nuanciertes verrständnins für autonomy in design
Overall, autonomy and agency-driven design have been a longstanding element in HCI~\cite{friedman_software_1997}. However, a lack of fine-grained understanding of how to connect these principles with usability~\cite{leimstadtner_investigating_2023} suggests more work is needed to take full advantage of existing design paradigms. Our model invites researchers and practitioners to reflect on their responsibilities. We therefore hope that it can support continuing work and foster more responsible designs. 
% - future work: deploy the model alongside design methods
Future work could study the effectiveness of the model alongside existing design methods. For example, it would be interesting to learn whether persuasive design or nudges can afford more autonomy if the model is used alongside such practices.
% - alternatives to user protection 
% - balancing autonomy
Nevertheless, existing work in HCI echoes a balancing act between user autonomy and alternation of choice architecture in every design. \textbf{It remains within the designers' responsibility to understand this challenge and follow ethical design principles promoting user autonomy.} A future goal of this work could be the investigation of opportune design strategies that better connect usability, user experience, and autonomy and create better incentives and tangible examples for practitioners to follow such principles.
% 1-2 implication statements in bold

\subsection{Ways Forward}
% Limitations and Future Work
Our proposed model mainly targets its depicted challenges from an HCI perspective. While the model offers novel insights in this regard and is designed to assist practitioners in their work, it is only one step towards understanding the underlying mechanisms of dark patterns. P15, a participant with a psychology background, speculated into the direction of cognitive processing:
\begin{quote}
    \textit{``I can come up with the whole information processing pipeline of human psychology here and just identify some [showing that] it's not covering everything.'' -- P15}
\end{quote}
Although the discussions of our participants did not come to an overall conclusion or identify specific solutions, they provided rich food for thought about the different angles from which dark patterns could be studied. With more emphasis on human cognition, perhaps future work can work toward models that better support users' perspectives. An already existing angle in this regard is presented in cognitive bias modification (CBM). The work highlighting the effectiveness in which autonomous decisions can be altered to be better reflected~\cite{hertel_cognitive_2011, jones_cognitive_2017} spotlights an exciting direction to help users make better decisions when faced with dark patterns. Moreover, research in HCI could adopt CBM to devise technologies that prepare and shape users' expectations in line with possible interactions.

% - Cognitive Bias Modification (CBM) could be used in design to directly change automatic cognitive processes

%%%%%%%%%%%%%%%%%%%%%%%%%%%%%
% LIMITATION & FUTURE WORK  %
%%%%%%%%%%%%%%%%%%%%%%%%%%%%%
\section{Limitations}
Although we were careful in designing and conducting this study, we recognise that this work is prone to certain limitations. Firstly, the last focus group was attended by only three instead of four participants (two experts in dark pattern scholarship and one expert in psychology). While the focus group functioned well in terms of interpersonal dynamics, a fourth member might have changed the dynamic of the discussions. Moreover, this changed the dynamic in which prior card sorting tasks were conducted. Instead of two groups including one expert from the respective fields to discuss and execute the task, we decided to have one group including all three participants. Again, the outcome was comparable to prior focus groups, but we cannot know whether a second expert with expertise in cognitive science or psychology would have impacted the group's discourse and, thus, our findings.

Secondly, the study was designed and administered by three researchers with backgrounds in HCI research. Although one of them has a background in psychology, their current research is situated in the field of HCI. Although the study was informed by relevant work on the concepts of cognitive biases and human behaviour to ensure a levelled discussion in terms of the considered subjects, we acknowledge a certain bias stemming from our expertise. We would welcome any future attempts to reproduce our findings in the fields of human cognition and behaviour to gain different views that may result in additional findings.

Our personal HCI backgrounds also resulted in a third limitation regarding the \textit{Relationship Model of Cognitive Biases and Dark Patterns}. Here, we recognise that our HCI background guided our moderation of the focus groups and informed our interpretation of the data as well as any further analysis. To mitigate these effects, we extensively reviewed material from the field of human cognition, particularly focusing on cognitive bias literature. However, researchers with other backgrounds may interpret the same data differently.

%%%%%%%%%%%%%%%%%%%%%%%%%%%%%
%        CONCLUSION         %
%%%%%%%%%%%%%%%%%%%%%%%%%%%%%
\section{Conclusion}
This research sheds light on the dynamic relationship between cognitive biases and dark patterns. Based on a focus group study with expert participants, we explore the ethical considerations throughout the development of design and describe the \textit{Relationship Model of Cognitive Biases and Dark Patterns} providing an overview of the continuum between design and real-world implications of harmful design.

To that end, we emphasize the critical role of practitioners and researchers in considering the ethical implications of their design decisions, particularly when it comes to user autonomy. By illustrating the dynamic process through which cognitive biases are leveraged to create dark patterns, the model underscores the responsibility designers hold in shaping user experiences. Furthermore, the study recognises the importance of safeguarding strategies for end-users and regulatory measures to protect individuals from the potential harms of dark patterns. This recognition aligns with broader discussions in the HCI community regarding better implementation of ethical design practices and user-centred development of autonomy and agency-enabling technologies.

In summary, this research deepens our understanding of the intricate interplay between cognitive biases and dark patterns but also provides a practical model for practitioners to navigate this complex landscape responsibly. As technology continues to play an increasingly ubiquitous role in our lives, the ethical considerations highlighted in this study become even more pertinent as related work catalogues the variety of unethical, dark patterns. Thus, we hope that this research contributes to the ongoing dialogue on ethical design in HCI and sets the stage for future investigations into creating safer, more user-friendly digital interfaces.
\bibliographystyle{ACM-Reference-Format}
\bibliography{references.bib}

\appendix

\end{document}

%% file: tables/participants.tex
\begin{table*}[t]
\begin{tabular}{@{}lllllll@{}}
\toprule
\multicolumn{7}{c}{\textbf{\textsc{Participant Table}}} \\ \midrule
ID & Age & Gender & Country of Residence & Education & Occupation & Years of Experience \\ \midrule
P1 & 47 & Male & Ireland & PhD & Senior UX Researcher & 3 \\
P2 & 32 & Male & Germany & PhD & Postdoc & 8 \\
P3 & 40 & Female & Netherlands & PhD & Assistant Professor & 7 \\
P4 & 25 & Male & United States & MSc & PhD Candidate & 3 \\
P5 & 28 & Female & United States & MSc & PhD Candidate & 5 \\
P6 & 28 & Male & Finland & PhD & Postdoc & 5 \\
P7 & 29 & Female & United States & MSc & PhD Candidate & 6 \\
P8 & 29 & Female & Germany & Diploma & PhD Candidate & 4 \\
P9 & 39 & Non-binary & United States & PhD & Professor & 15 \\
P10 & 31 & Male & Ireland & PhD & Postdoc & 6 \\
P11 & 32 & Male & Germany & PhD & Postdoc & 7 \\
P12 & 40 & Male & Ireland & PhD & Assistant Professor & 10 \\
P13 & 29 & Female & Luxembourg & MSc & PhD Candidate & 3 \\
P14 & 29 & Female & United States & PhD & Assistant Professor & 12 \\
P15 & 35 & Male & Finland & PhD & Assistant Professor & 6 \\ \midrule
\multicolumn{1}{r}{\textit{Mean}} & $=32.87$ & & & & \multicolumn{1}{r}{\textit{Mean}} & $=6.73$\\
\multicolumn{1}{r}{\textit{SD}} & $=6.08$ & & & & \multicolumn{1}{r}{\textit{SD}} & $=3.43$\\
\bottomrule
\end{tabular}

\caption{This table presents an overview of expert participants of the four focus groups.}
\label{tab:participants}
\end{table*}